\documentclass[12pt,preprint]{aastex}
\usepackage{natbib}
\usepackage{amsmath}
\usepackage{rotating}

\begin{document}

\title{synthetic off-axis light curves for low energy gamma-ray bursts}
\author{Hendrik J. van Eerten, Andrew I. MacFadyen}
\affil{
  Center for Cosmology and Particle Physics, Physics
  Department, New York University, New York, NY 10003}

\begin{abstract}
We present results for a large number of gamma-ray burst (GRB) afterglow light curve calculations, done by combining high resolution two-dimensional relativistic hydrodynamics simulations using \textsc{ram} with a synchrotron radiation code. Results were obtained for jet energies, circumburst medium densities and jet angles typical for short and underluminous GRBs, different observer angles and observer frequencies from low radio (75 MHz) to X-ray (1.5 keV). We summarize the light curves through smooth power law fits with up to three breaks, covering jet breaks for small observer angles, the rising phase for large observer angles and the rise and decay of the counterjet. All light curve data are publicly available via \url{http://cosmo.nyu.edu/afterglowlibrary}. The data can be used for model fits to observational data and as an aid for predicting observations by future telescopes such as LOFAR or SKA and will benefit the study of neutron star mergers using different channels, such as gravitational wave observations with LIGO or Virgo.

For small observer angles, we find jet break times that vary significantly between frequencies, with the break time in the radio substantially postponed. Increasing the observer angle also postpones the measured jet break time. The rising phase of the light curve for large observer angle has a complex shape that can not always be summarized by a simple power law. Except for very large observer angles, the counter jet is a distinct feature in the light curve, although in practice the signal will be exceedingly difficult to observe by then.
\end{abstract}

\section{Introduction}

Short gamma-ray bursts (SGRBs) are likely produced by neutron star-neutron star or neutron star-black hole mergers (see e.g. \citealt{Eichler1989} for an early exploration of this idea). This makes them physically different from long duration GRBs, which result from the stellar collapse of a massive star. The distribution of GRB durations is therefore expected (and found) to be bimodal rather than continuous \citep{Kouveliotou1993}. Nevertheless, theoretical models of both types of GRB share many similarities, the most important of them being the formation of an ultrarelativistic jet. SGRB jet models generally probe a different part of the possible parameter space for such jets than long GRBs, although some overlap exists with underluminous instances of the latter (e.g. GRB 100316D, \citealt{Starling2011}). The overall energy release for SGRBs is of the order $10^{48-50}$ ergs (rather than $10^{52}$ ergs), the circumburst particle densities of order $10^{-5} - 1$ cm$^{-3}$ (rather than 1 cm$^{-3}$) and they are less collimated (although there is currently little observational confirmation of the latter, in hydrodynamical models for long GRBs, e.g. \citealt{MacFadyen1999}, the jet becomes collimated by passing through a dense stellar interior. This mechanism is absent for SGRBs). Reviews of SGRB science can be found in \cite{Nakar2007}, \cite{Gehrels2009} and a recent comparison to long GRBs in \cite{Nysewander2009}.

Like long GRBs, short and underluminous GRBs produce afterglows peaking at progressively longer wavelengths with time, although they are harder to detect because they are intrinsically fainter. Analytical models of SGRB afterglows suffer from the same simplifications and shortcomings as those of long GRBs (mainly with respect to jet decollimation and off-axis emission). These can be addressed through combining high-resolution relativistic hydrodynamics (RHD) simulations with numerical radiative transfer for synchrotron radiation. Such simulations have already been performed for long GRBs (e.g. \citealt{Zhang2009, vanEerten2010}). A more accurate understanding of short GRB afterglows is currently especially interesting not only because they are actually being detected starting a few years ago, but also because a new generation of extremely sensitive detectors of SGRBs is becoming operational. On the one hand there are instruments such as LOFAR or SKA, that will detect SGRB (afterglow) emission through the traditional electromagnetic (EM) channel, but at unprecedented long wavelengths on the order of tens of MHz rather than GHz (The lower limit goal for SKA is 60 MHz, for SKA pioneer project ASKAP it is 300 MHz, \citealt{Johnston2008}. for LOFAR it is $\sim 10$ MHz, \citealt{Rottgering2006}). Transient monitoring campaigns with these instruments should be able to detect afterglows even if the prompt emission remains unseen. On the other hand, completely new channels are becoming available for GRB detection: multiple gravitational-wave (GW) detectors are currently in operation (e.g. LIGO, \citealt{Abbott2009} and Virgo, \citealt{Acernese2008}) and upgrades are anticipated. The amount of information that can be obtained from GW detections can be significantly enhanced by information from their EM counterparts that can help break degeneracies in GW model fits \citep{Nissanke2010}. Also, the expected observer time between the GW signal and the peak of the afterglow signal is expected to be on the order of several days at least, so a GW localization can be used to increase the odds of detecting an afterglow. It is therefore important to accurately understand the relationships between SGRB energy, collimation and observer angle and their observational implications (see also \citealt{Nakar2011}).

To further our understanding of SGRB afterglows we have performed a series of two-dimensional RHD simulations of SGRB jets interacting with the circumburst environment. In this letter we present, for the first time, short GRB afterglow light curves  for observers positioned both on the jet axis and off-axis that are calculated from RHD simulations and include both synchrotron emission and synchrotron self-absorption. In section \ref{simulations_section} we explain our methods. In section \ref{typical_case_section} we discuss light curves and spectra for a number of cases. All data are publicly available at \url{http://cosmo.nyu.edu/afterglowlibrary}, but in addition to this we provide tables showing results for smoothly broken power law fits to the light curves for the cases under discussion. The fits should be taken as a brief summary of the overall shapes of the light curves, rather than a definitive description that completely captures the underlying physics. In section \ref{discussion_section} we draw conclusions.

\section{Simulations and radiation}
\label{simulations_section}

We have performed 12 RHD simulations in 2D spherical coordinates using the parallel \textsc{ram} adaptive-mesh refinement code \citep{Zhang2006}, four of which we will describe in detail in this paper (the resulting light curves for the others are publicly available on the website). Taking a conic section from the Blandford-McKee (BM) self-similar solution for a relativistic explosion in a homogeneous medium \citep{Blandford1976} as starting point, these simulations cover all possible combinations of the following values for the physical quantities determining the dynamics of the system: energy in both jets $E_j = 10^{48}$ or $10^{50}$ ergs, circumburst number density $n = 10^{-5}$, $10^{-3}$, $1$ cm$^{-3}$, jet half opening angle $\theta_j = 0.2$, $0.4$ rad. Numerical results for the dynamics of spreading and decelerating BM jets have been described in detail in \cite{Zhang2009}. Based on these findings, we have chosen a starting BM jet fluid Lorentz factor directly behind the shock front of $\gamma = 10$ (well in excess of $1 / \theta_j$, both for $\theta_j = 0.2$ and $0.4$, when lateral spreading is expected to set in), a lab frame stopping time $t = 10 t_{NR}$ and corresponding grid size $r = c t$ (where $c$ the speed of light and $t_{NR}$ as defined below) and a grid resolution such that the initial blast wave width $R / \Gamma^2$ (with $R$ the blast wave radius and $\Gamma$ the shock Lorentz factor) was resolved by approximately 100 grid cells. In practice, the latter requirement lead us to 24 base level blocks (of 16 cells each) in the radial direction and 12 initial levels of refinement. We have used 2 base level blocks in the angular direction. As in earlier work, we gradually decreased the peak refinement level over time. In addition, we kept the peak refinement level for the inner regions of the jet a few levels lower than that of the outer regions. This avoids spending too much computational effort on resolving Kelvin-Helmholtz type instabilities in the flow that have little effect on the radiation and overall dynamics. The time $t_{NR}$ reflects the time when the blast wave was analytically expected \citep{Piran2005} to become nonrelativistic and settle into the self-similar Sedov-Taylor solution \citep{Taylor1950, Sedov1959} and is given by
\begin{equation}
 t_{NR} \approx 970 E^{1/3}_{iso,53} n^{-1/3} \textrm{ days.}
\end{equation}
Here $E_{iso,53}$ denotes isotropic equivalent energy in units of $10^{53}$ ergs, related to the energy in both jets according to $E_{iso} = 2 E_j / \theta_j^2$. The theoretical value for $t_{NR}$ was numerically found to underestimate the transition time and duration to spherical nonrelativistic flow and \cite{Zhang2009} found that the transition time is better approximated by $\backsim 5 t_{NR}$.

We have combined simulation output (3000 snapshots per simulation) with a linear radiative transfer code that calculates the observed flux at various observer angles from rays through the evolving fluid, including synchrotron emission and synchrotron self-absorption. The current approach generalizes the radiation code described in \cite{vanEerten2009, vanEerten2010} to off-axis observers. It further differs from these studies in that it follows the simplified approach to the general synchrotron emission used in \cite{vanEerten2010c} that uses a global approach to electron cooling rather than a local one and that in turn has been based on \cite{Sari1998}. The total number of rays required for a single observation is calculated through a procedure analogous to adaptive mesh refinement, where each possible refinement doubles the number of rays in a single group (or ``block'', containing 16 rays) either in the $r$ or $\phi$ direction, where $r$ and $\phi$ are polar coordinates on the plane perpendicular to the direction of the rays (i.e. towards the observer). A local total of 9 refinements (in any combination of radial and angular refinements), starting with 24 base level blocks in the radial direction ($r = 0$ - fluid grid maximum) and 2 in the angular direction ($\phi = 0 - \pi$)  was found to be sufficient to converge on a fixed flux value.  Although our method provides us with spatially resolved images for off-axis observers as well as fluxes, these will be presented elsewhere.

As usual in afterglow modeling, a number of parameters is used to capture the radiation physics. The accelerated electron power law slope has been set to $p = 2.5$, the fraction of accelerated electrons $\xi_N = 1.0$, the energy in these electrons as fraction of the thermal energy $\epsilon_E = 0.1$, the fraction of thermal energy in the magnetic field $\epsilon_B = 0.1$.

In order to ensure complete coverage at early observer times we have used an analytical implementation of the BM solution at fluid Lorentz factors $> 10$ rather than simulation output to calculate emission and absorption. Early time contributions have been confirmed to connect smoothly to those from simulation output. However, even with additional early time contribution, the earliest observer time with effectively full coverage still differs between observer angles.

\section{Light curves and spectra}
\label{typical_case_section}

We haved calculated light curves at the following four frequencies: 75 MHz (radio, \textsc{LOFAR, SKA}), 1.43 GHz (radio, \textsc{WSRT, VLA}), $4.56 \times 10^{14}$ Hz (R-band, \textsc{VLT}) and $3.63 \times 10^{17}$ Hz ($1.5$ KeV X-rays, \emph{Swift} \textsc{XRT}). In this letter we summarize and discuss the results in detail for the following cases:
\begin{itemize}
 \item $E_j = 10^{48}$ ergs, $\theta_j = 0.2$ rad, $n = 10^{-3}$ cm$^{-3}$ (A)
 \item $E_j = 10^{48}$ ergs, $\theta_j = 0.4$ rad, $n = 10^{-3}$ cm$^{-3}$ (B)
 \item $E_j = 10^{50}$ ergs, $\theta_j = 0.4$ rad, $n = 10^{-3}$ cm$^{-3}$ (C)
 \item $E_j = 10^{50}$ ergs, $\theta_j = 0.4$ rad, $n = 1$ cm$^{-3}$ (D)
\end{itemize}
This way we cover both small and large opening angles and the effect of increased jet energy and circumburst density. For all cases we have computed light curves for a range of observer angles: $\theta_{obs} = 0$, 0.1, 0.2, 0.4, 0.8 and $\pi / 2$ rad.

\begin{figure}[h]
 \centering
 \includegraphics[width=0.6\columnwidth]{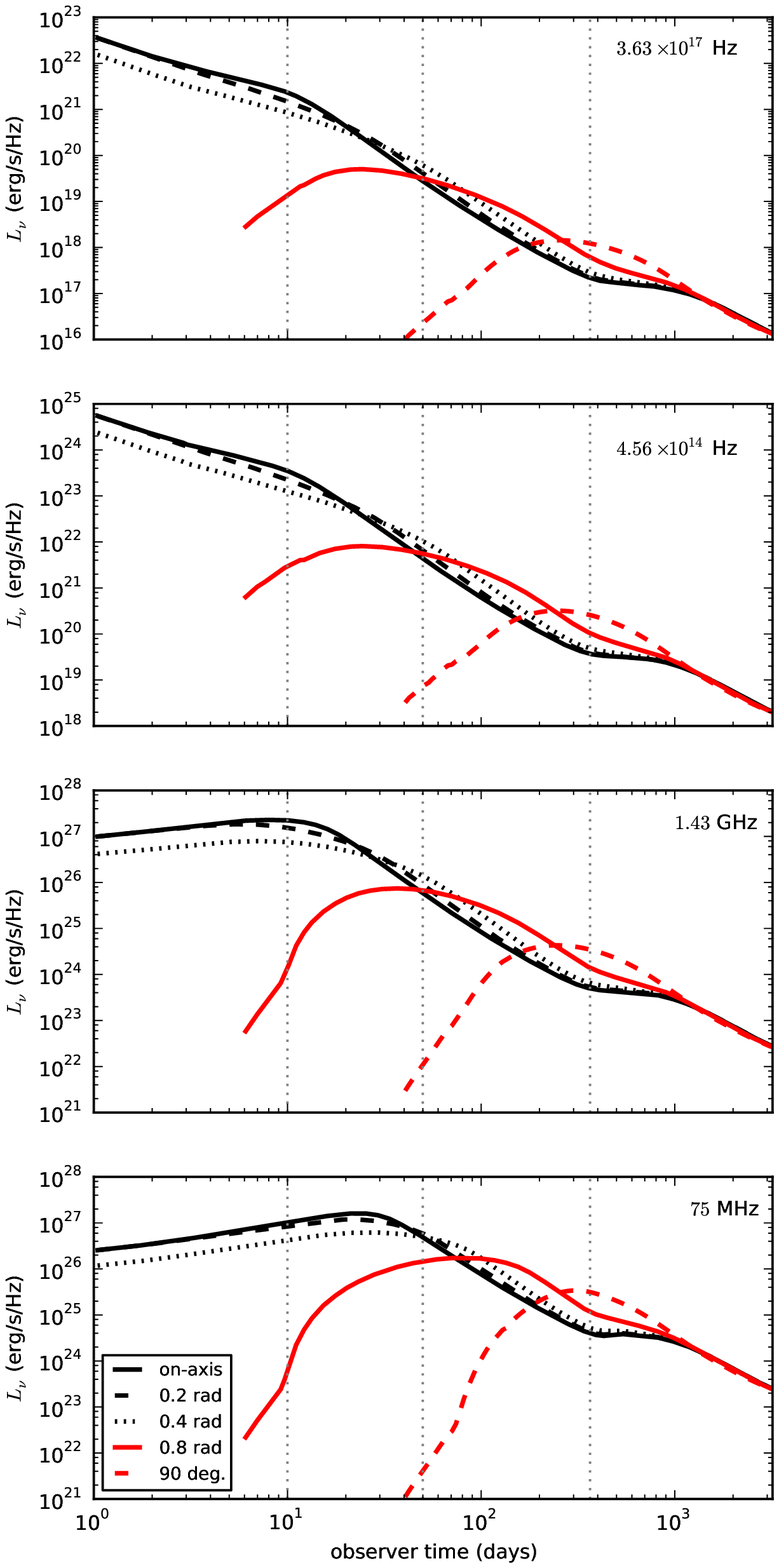}
 \caption{Observed luminosity light curves for $E_j = 10^{48}$ ergs, $\theta_j = 0.4$ rad, $n = 10^{-3}$ cm$^{-3}$ (case B). Observer frequencies from top to bottom: $3.63 \times 10^{17}$ Hz, $4.56 \times 10^{14}$ Hz, 1.43 GHz and 75 MHz. The legend applies to all plots. 10 days, 50 days and 1 yr have been marked with vertical dotted grey lines. Spectra for these times are provided in Fig. \ref{spectra_figure}.}
 \label{lightcurves_figure}
\end{figure}
\begin{figure}[h]
 \centering
 \includegraphics[width=0.6\columnwidth]{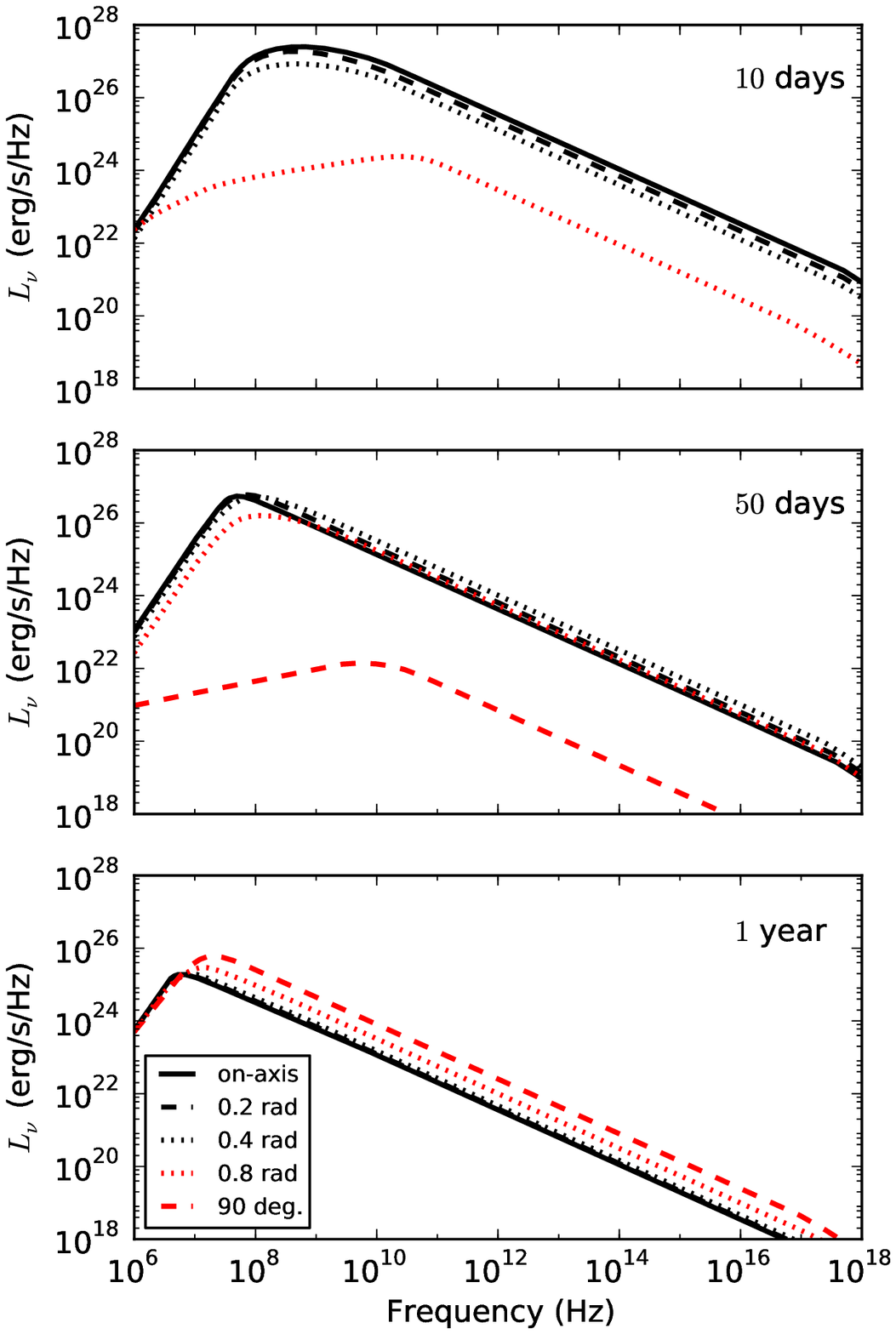}
 \caption{Spectra for $E_j = 10^{48}$ ergs, $\theta_j = 0.2$ rad, $n = 10^{-3}$ cm$^{-3}$ at $t_{obs} = 10$ days, $50$ days and 1 yr (top to bottom plot), for various observer angles. The legend applies to all plots.}
 \label{spectra_figure}
\end{figure}

The resulting light curves for case B have been plotted in Fig. \ref{lightcurves_figure} For three observer times we have calculated broadband spectra as well, and these are plotted in Fig. \ref{spectra_figure}. For the whole set of simulations (case A-D), we have summarized the shapes of the light curves by fitting smoothly connected power laws in time, using
\begin{eqnarray}
 F & = & F_0 \left[ \left(\frac{t}{t_{01}} \right)^{-s_{01} \beta_0} + \left(\frac{t}{t_{01}} \right)^{-s_{01} \beta_1} \right]^{-1/s_{01}} \times \nonumber \\
 & & \left[ 1 + \left( \frac{t}{t_{12}} \right)^{s_{12}(\beta_1 - \beta_2)} \right]^{-\frac{1}{s_{12}}} \left[ 1 + \left( \frac{t}{t_{23}} \right)^{s_{23}(\beta_2 - \beta_3)} \right]^{-\frac{1}{s_{23}}},
\end{eqnarray}
for all observer angles except $\pi / 2$, when both jets are seen exactly on edge and
\begin{eqnarray}
 F & = & F_0 \left[ \left(\frac{t}{t_{01}} \right)^{-s_{01} \beta_0} + \left(\frac{t}{t_{01}} \right)^{-s_{01} \beta_1} \right]^{-1/s_{01}},  
\end{eqnarray}
is sufficient. Using multiple smoothly connected power laws to describe the data is common both in theoretical and observational studies (e.g. \citealt{, Beuermann1999, Granot2002}). Fit parameter $F_0$ sets the scale of the light curve (here, we have set redshift $z=0$ and observer luminosity distance $d_L = 10^{28}$ cm). Different power law regimes meet at transition points $t_{01}$, $t_{12}$ and $t_{23}$, measured in days. The slopes at the different regimes are given by $\beta_0$, $\beta_1$ and $\beta_2$, while the sharpnesses for the transitions are given by $s_{01}$, $s_{12}$, $s_{23}$.

Fit results for case A-D are listed in tables \ref{characteristics_table} and \ref{characteristics2_table}. Complete datasets are publicly available (at \url{http://cosmo.nyu.edu/afterglowlibrary}) and we emphasize that the power law fits are meant only as convenient summary, rather than as a full description based on the underlying physics of the afterglows. Nevertheless, the first temporal break $t_{01}$ can be roughly interpreted as the jet break time for small observer angles. For high observer angles it marks the difference between the rise (where relativistic beaming dominates the shape of the light curve, e.g. \citealt{Granot_etal_2002_ApJ}) and decay of the signal. The second and third break $t_{12}$, $t_{23}$ are used to summarize the rise and decline of the counterjet.

\begin{figure}[h]
 \centering
 \includegraphics[width=0.6\columnwidth]{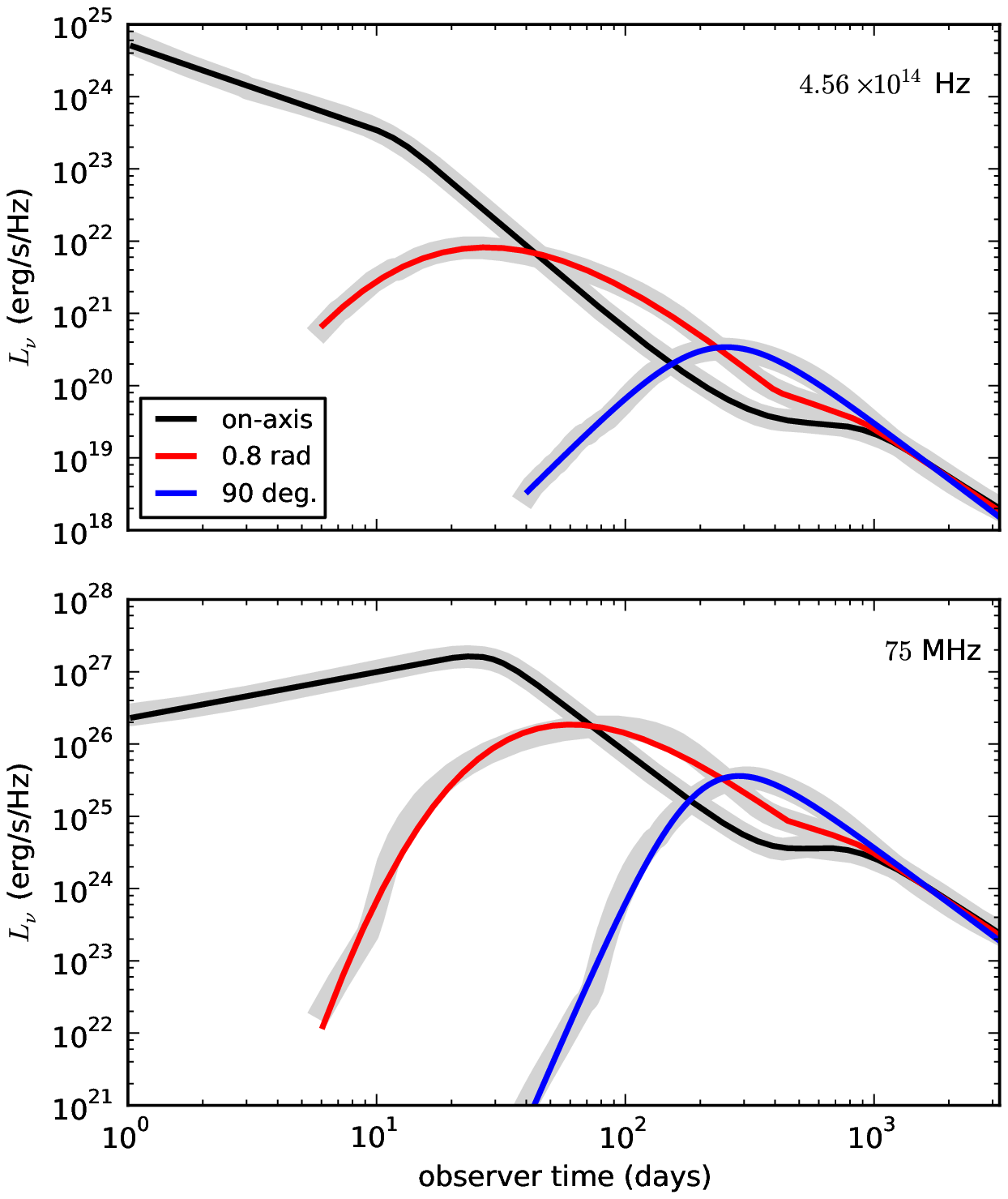}
 \caption{Direct comparison between power law fits and simulated light curves (broad light grey curves) for case B. The legend applies to both plots. The top plot shows optical ($4.56 \cdot 10^{14}$ Hz) light curves, the bottom plot shows low radio ($75 \cdot 10^{6}$ Hz) light curves.}
 \label{fits_vs_lightcurves_figure}
\end{figure}

Fitting for 11 fit parameters allows for a lot of freedom, and in most cases the resulting fit function captures the simulation light curve with only an occasional difference of up to a few percent. In Fig. \ref{fits_vs_lightcurves_figure} we have plotted a number of fit results for case B, that illustrate the accuracy of the power law fits. The fit results have been obtained using a straighforward implementation of nonlinear least squares fitting while assuming a fractional error of 10 percent on the simulation points. The number of datapoints varies between $\sim$ 200 for on-axis observers and $\sim$ 100 for observers completely off axis. The number of degrees of freedom (dof) varies similarly, with 11 fit parameters (or 5, for a single break) subtracted from the number of datapoints. All fits had $\chi^2 / \text{dof} < 1$, except for a number of separately marked cases. Sometimes fits were insensitive to some of the fit parameters. Where this was the case (when the error on the fit parameter was comparable in size to the parameter itself), only a single digit is given in the table.

From Figs. \ref{lightcurves_figure}, \ref{spectra_figure} and the tabulated results, we draw conclusions regarding the general structure of SGRB light curves. For small observer angles, the results clearly confirm that afterglow jet breaks are \emph{chromatic}, which was first reported in \cite{vanEerten2010b} (albeit there for long GRB's, both for 1D top hat jet simulations and a medium-resolution 2D simulation using a different hydrodynamics code). At low (radio) frequencies, the jet break is consistently postponed with respect to the jet break in the optical and X-ray. Although the fitted break times also differ between optical and X-ray, the current approach that approximates the electron cooling time by the explosion duration is not sufficiently realistic to allow for definitive statements on the chromaticity of the break between the two frequencies (a detailed treatment of electron cooling lies beyond the scope of this work. The practical relevance of off-axis X-ray afterglow light curves for SGRBs is limited). Although we have not accounted for this in our fits, the jet break splits into two breaks when the observer moves off-axis. The $\theta_{obs} = 0.2$ rad light curves from Fig. \ref{lightcurves_figure} provide an example of this: at early times they have again joined their repective on-axis light curves. For observers exactly on edge, only the late time break is left, with the early time break having moved to $t = 0$. This wide range of jet breaks is unfortunate from a practical point of view. It means that an equation for the break time $t_j$ such as
\begin{equation}
t_j = 3.5 (1+z) E_{iso,53}^{1/3} n_1^{-1/3} \left(\frac{\theta_j +
\theta_{obs}}{0.2}\right)^{8/3} \textrm{ days},
\label{jetbreak_time_equation}
\end{equation}
from \cite{vanEerten2010c} will be consistent with the data (as can be seen from a comparison to values in tables \ref{characteristics_table} and \ref{characteristics2_table}), but only given the wide range of jet break times across the frequencies. Nevertheless, the scalings in eq. \ref{jetbreak_time_equation} are confirmed.

All light curves confirm the clear rise of the counterjet that was both analytically expected (e.g. \citealt{Granot2003}) and a robust feature of earlier numerical work \citep{Zhang2009}. They show that at late times and for off-axis observers, synchrotron self-absorption (s.s.a.) has little effect on the light curves. For case B, this can be seen directly in Fig. \ref{spectra_figure}. For small opening angles and radio frequencies, the influences of s.s.a. and the jet opening angle on the light curve can become hard to disentangle. The most extreme case is provided by case D, where the circumburst density $n$ is the highest. This is not unexpected given that the synchrotron break frequency $\nu_{sa}$ scales as $\nu_{sa} \propto n^{3/5}$ when $\nu_{sa}$ lies below the synchrotron break frequency $\nu_m$ and as $\nu_{sa} \propto n^{4/13}$ otherwise \citep{Granot2002}.

Light curves for large observer angles exhibit a steep rising phase that will be more complex than a straight power law, especially at low frequencies. The low radio light curve plots in Fig \ref{lightcurves_figure} for $\theta_{obs}$ equal to 0.8 and $\pi/2$ provide examples. This was not reported in earlier numerical studies (e.g. \citealt{vanEerten2010c, Granot2001}, where the early rising part has been truncated from the light curve). It is a genuine feature that depends on the spectral shape  rather than lateral spreading and is also seen in simplified analytical models \citep{Rhoads1999, Sari1998, Granot_etal_2002_ApJ}).

A long term feature common to all light curves not fully captured by the power law fits is the gradual transition from relativistic to nonrelativistic flow. At the last observer time covered the simulations have not yet been nonrelativistic sufficiently long for the expected nonrelativistic slope above $\nu_m$ of $\beta = -1.65$ to become dominant (note that at any given observer time the observed flux is the combined signal from a range of emission times). In practice, observing a SGRB afterglow fully in the nonrelativistic regime will be exceedingly unlikely.

\section{Discussion and conclusions}
\label{discussion_section}

We have performed a series of high-resolution RHD simulations in 2D to calculate the jet outflow for physical parameters typical of those expected for subenergetic GRB's. From these we have calculated afterglow light curves at various frequencies, covering low radio (75 MHz) up to X-ray (1.5 keV) and for observer angles from 0 to $\pi/2$ rad. The data for all light curves from this paper are publicly available via \url{http://cosmo.nyu.edu/afterglowlibrary}, that also provides results from a more extensive probe of parameter space. We summarize the light curves via smooth power law fits that capture features such as the jet break for small observer angles, the early time rise due to relativistic beaming for high observer angles and the rise and decay of the counterjet. The results here present the most accurate calculations to date of light curve predictions of the standard afterglow jet theory as it applies to short GRB's, fully accounting for aspects such as jet spreading, observer position and arrival time effects.  Although we do not discuss this in detail in this work, the light curves show that SGRB / underluminous afterglows should in principle be observable (at least in the radio) even for observers outside the jet cone. The light curves in this paper and in the on-line database should prove useful for detectability estimates using future radio telescopes such as SKA and LOFAR. Such estimates will also benefit the gravitational waves community, since the amount of information that can be extracted from GW measurements increases significantly when EM counterparts are observed as well. Furthermore, GW observations can aid the search for EM counterparts.

In general, the afterglow light curves are well described by smooth power law fits with up to three breaks, although sometimes the rising phase for high observer angles is problematic. The effects of increasing circumburst density and jet energy are as expected from theoretical models. The jet break for small observer angles varies greatly between frequencies, confirming a result from \cite{vanEerten2010b}. Increasing the observer angle postpones the jet break (it really splits the break into two separate breaks, but the second break is the strongest).

\section{Acknowledgements}
This work was supported in part by NASA under Grant No. 09-ATP09-0190 issued through the Astrophysics Theory Program (ATP).  The software used in this work was in part developed by the DOE-supported ASCI/Alliance Center for Astrophysical Thermonuclear Flashes at the University of Chicago.

%\bibliography{sgrbs}

\begin{table}
\tiny{
\centering
\begin{tabular}{|ll||ll|lll|lll|lll|}
\hline
 $\theta_{obs}$ & $\nu$ (Hz) & $F_0$ & $\beta_0$ & $t_{01}$ & $s_{01}$ & $\beta_1$ & $t_{12}$ & $s_{12}$ & $\beta_2$ & $t_{23}$ & $s_{23}$ & $\beta_3$ \\
\hline
 $0.0$ & $75 \cdot 10^{6}$ & $1.6 \cdot 10^{-4}$ & $0.31$ & $28.0$ & $2.7$ & $-2.6$ & $9 \cdot 10^{2}$ & $-2 \cdot 10^{-2}$ & $2 \cdot 10^{2}$ & $9 \cdot 10^{2}$ & $2 \cdot 10^{-2}$ & $-2.6$ \\
 $0.0$ & $1.43 \cdot 10^{9}$ & $4.9 \cdot 10^{-4}$ & $0.24$ & $10.0$ & $1.2$ & $-3.0$ & $1.1 \cdot 10^{3}$ & $-0.15$ & $8.9$ & $1.3 \cdot 10^{3}$ & $0.49$ & $-1$ \\
 $0.0$ & $4.56 \cdot 10^{14}$ & $1 \cdot 10^{-6}$ & $-1.2$ & $3.5$ & $0.66$ & $-3.0$ & $1.1 \cdot 10^{3}$ & $-0.15$ & $8.6$ & $1.3 \cdot 10^{3}$ & $0.52$ & $-1$ \\
 $0.0$c & $3.63 \cdot 10^{17}$  & $2.3 \cdot 10^{-9}$ & $-1.5$ & $4.5$ & $1.5$ & $-3.0$ & $1.9 \cdot 10^{3}$ & $-0.10$ & $9.6$ & $1.6 \cdot 10^{3}$ & $0.77$ & $2$ \\

 $0.1$ & $75 \cdot 10^{6}$  & $1.4 \cdot 10^{-4}$ & $0.34$ & $31.0$ & $2.3$ & $-2.7$ & $9 \cdot 10^{2}$ & $-2 \cdot 10^{-2}$ & $2 \cdot 10^{2}$ & $10 \cdot 10^{2}$ & $2 \cdot 10^{-2}$ & $-2.6$ \\
 $0.1$ & $1.43 \cdot 10^{9}$  & $3.5 \cdot 10^{-4}$ & $0.15$ & $13.0$ & $1.4$ & $-3.1$ & $1.1 \cdot 10^{3}$ & $-0.15$ & $8.9$ & $1.3 \cdot 10^{3}$ & $0.49$ & $-1$ \\
 $0.1$ & $4.56 \cdot 10^{14}$  & $4.4 \cdot 10^{-8}$ & $-1.8$ & $11.0$ & $4.8$ & $-3.1$ & $1.1 \cdot 10^{3}$ & $-0.15$ & $8.8$ & $1.3 \cdot 10^{3}$ & $0.50$ & $-1$ \\
 $0.1$c & $3.63 \cdot 10^{17}$ & $3.1 \cdot 10^{-10}$ & $-1.7$ & $9.7$ & $6.8$ & $-3.1$ & $2.1 \cdot 10^{3}$ & $-9.3 \cdot 10^{-2}$ & $10.0$ & $1.6 \cdot 10^{3}$ & $0.77$ & $2$ \\

 $0.2$ & $75 \cdot 10^{6}$ & $1.2 \cdot 10^{-4}$ & $0.48$ & $38.0$ & $1.7$ & $-2.8$ & $10 \cdot 10^{2}$ & $-2 \cdot 10^{-2}$ & $2 \cdot 10^{2}$ & $10 \cdot 10^{2}$ & $2 \cdot 10^{-2}$ & $-2.5$ \\
 $0.2$ & $1.43 \cdot 10^{9}$ & $2.3 \cdot 10^{-4}$ & $0.27$ & $1.9 \cdot 10^{1}$ & $1.1$ & $-3.4$ & $1.0 \cdot 10^{3}$ & $-0.13$ & $8.9$ & $1.3 \cdot 10^{3}$ & $0.50$ & $-0.9$ \\
 $0.2$ & $4.56 \cdot 10^{14}$ & $1.9 \cdot 10^{-8}$ & $-1.5$ & $18.0$ & $4.3$ & $-3.3$ & $1.0 \cdot 10^{3}$ & $-0.14$ & $9.3$ & $1.3 \cdot 10^{3}$ & $0.46$ & $-1$ \\
 $0.2$c & $3.63 \cdot 10^{17}$ & $1.3 \cdot 10^{-10}$ & $-1.5$ & $16.0$ & $5.3$ & $-3.3$ & $1.8 \cdot 10^{3}$ & $-0.10$ & $9.5$ & $1.5 \cdot 10^{3}$ & $0.73$ & $2$ \\

\hline

 $0.4$a & $75 \cdot 10^{6}$ & $4$ & $8.1$ & $14.0$ & $5.8 \cdot 10^{-2}$ & $-4.5$ & $4.9 \cdot 10^{2}$ & $-1$ & $0.8$ & $1.1 \cdot 10^{3}$ & $2$ & $-3.1$ \\
 $0.4$a & $1.43 \cdot 10^{9}$ & $7$ & $1.0 \cdot 10^{1}$ & $8.5$ & $5.5 \cdot 10^{-2}$ & $-5.1$ & $3.9 \cdot 10^{2}$ & $-0.55$ & $0.7$ & $1.2 \cdot 10^{3}$ & $2.4$ & $-2.9$ \\
 $0.4$a & $4.56 \cdot 10^{14}$ & $3 \cdot 10^{-4}$ & $3.8$ & $15.0$ & $7 \cdot 10^{-2}$ & $-5.6$ & $4.1 \cdot 10^{2}$ & $-0.45$ & $0.7$ & $1.2 \cdot 10^{3}$ & $2.1$ & $-3.1$ \\
 $0.4$a & $3.63 \cdot 10^{17}$ & $1 \cdot 10^{-6}$ & $5.1$ & $7.3$ & $7.4 \cdot 10^{-2}$ & $-4.6$ & $5.7 \cdot 10^{2}$ & $-0.3$ & $2$ & $1.3 \cdot 10^{3}$ & $1$ & $-2.6$ \\

 $0.8$a & $75 \cdot 10^{6}$ & $5.7 \cdot 10^{-6}$ & $7.5$ & $64.0$ & $4$ & $2.0$ & $4.1 \cdot 10^{2}$ & $-2 \cdot 10^{-2}$ & $2 \cdot 10^{2}$ & $4 \cdot 10^{2}$ & $2 \cdot 10^{-2}$ & $-3.1$ \\
 $0.8$a &  $4.56 \cdot 10^{14}$ & $2.9 \cdot 10^{-5}$ & $6.8$ & $77.0$ & $0.38$ & $-3.4$ & $4.8 \cdot 10^{2}$ & $-8$ & $-0.57$ & $1.1 \cdot 10^{3}$ & $7$ & $-2.8$ \\
 $0.8$ & $3.63 \cdot 10^{17}$ & $4.5 \cdot 10^{-9}$ & $3.7$ & $92.0$ & $0.28$ & $-4.3$ & $4.5 \cdot 10^{2}$ & $-2.8$ & $-0.58$ & $1.1 \cdot 10^{3}$ & $7.2$ & $-2.8$ \\
 $0.8$ & $3.63 \cdot 10^{17}$ & $7.4 \cdot 10^{-12}$ & $3.7$ & $80.0$ & $0.47$ & $-3.2$ & $5.3 \cdot 10^{2}$ & $-5.2$ & $-0.64$ & $1.3 \cdot 10^{3}$ & $9$ & $-2.7$ \\

 $\pi / 2$a & $75 \cdot 10^{6}$ & $6.5 \cdot 10^{-6}$ & $6.8$ & $4.1 \cdot 10^{2}$ & $0.75$ & $-2.9$ & & & & & & \\ 
 $\pi / 2$a & $1.43 \cdot 10^{-9}$ & $1.6 \cdot 10^{-6}$ & $6.1$ & $3.3 \cdot 10^{2}$ & $0.41$ & $-3.0$ & & & & & & \\
 $\pi / 2$a & $4.56 \cdot 10^{14}$ & $6.1 \cdot 10^{-11}$ & $3.5$ & $4.0 \cdot 10^{2}$ & $0.75$ & $-3.0$ & & & & & & \\
 $\pi / 2$a & $3.63 \cdot 10^{17}$ & $2.3 \cdot 10^{-13}$ & $3.5$ & $4.4 \cdot 10^{2}$ & $0.69$ & $-2.8$ & & & & & & \\

\hline
\hline
 $0.0$ & $75 \cdot 10^{6}$ & $1.6 \cdot 10^{-4}$ & $0.65$ & $29.0$ & $2.3$ & $-2.6$ & $5 \cdot 10^{2}$ & $-4 \cdot 10^{-2}$ & $90.0$ & $6 \cdot 10^{2}$ & $4 \cdot 10^{-2}$ & $-2.3$ \\
 $0.0$ & $1.43 \cdot 10^{9}$ & $2.5 \cdot 10^{-4}$ & $0.45$ & $13.0$ & $1.4$ & $-3.0$ & $6.3 \cdot 10^{2}$ & $-0.2$ & $5$ & $9.2 \cdot 10^{2}$ & $0.6$ & $-1.7$ \\
 $0.0$  & $4.56 \cdot 10^{14}$ & $2.2 \cdot 10^{-8}$ & $-1.2$ & $12.0$ & $5.3$ & $-3.0$ & $6 \cdot 10^{2}$ & $-0.2$ & $6$ & $8.8 \cdot 10^{2}$ & $0.5$ & $-1.8$ \\
 $0.0$ & $3.63 \cdot 10^{17}$  & $1.5 \cdot 10^{-10}$ & $-1.2$ & $12.0$ & $6.7$ & $-3.0$ & $8 \cdot 10^{2}$ & $-0.2$ & $6$ & $1.1 \cdot 10^{3}$ & $0.6$ & $-1$ \\
 
 $0.1$ & $75 \cdot 10^{6}$ & $1.6 \cdot 10^{-4}$ & $0.63$ & $31.0$ & $1.7$ & $-2.7$ & $5 \cdot 10^{2}$ & $-5 \cdot 10^{-2}$ & $80$ & $6 \cdot 10^{2}$ & $5 \cdot 10^{-2}$ & $-2.3$ \\
 $0.1$ & $1.43 \cdot 10^{9}$ & $2.8 \cdot 10^{-4}$ & $0.50$ & $14.0$ & $0.85$ & $-3.1$ & $6 \cdot 10^{2}$ & $-0.2$ & $5$ & $9.1 \cdot 10^{2}$ & $0.6$ & $-1.7$ \\
 $0.1$ & $4.56 \cdot 10^{14}$ & $1.6 \cdot 10^{-8}$ & $-1.2$ & $14.0$ & $2.4$ & $-3.0$ & $6 \cdot 10^{2}$ & $-0.2$ & $7$ & $8.6 \cdot 10^{2}$ & $0.4$ & $-1.9$ \\
 $0.1$ & $3.63 \cdot 10^{17}$  & $1.1 \cdot 10^{-10}$ & $-1.2$ & $14.0$ & $2.9$ & $-3.0$ & $8 \cdot 10^{2}$ & $-0.2$ & $6$ & $1.1 \cdot 10^{3}$ & $0.5$ & $-2$ \\

 $0.2$ & $75 \cdot 10^{6}$ & $1.4 \cdot 10^{-4}$ & $0.56$ & $37.0$ & $1.1$ & $-2.9$ & $5 \cdot 10^{2}$ & $-4 \cdot 10^{-2}$ & $80$ & $6 \cdot 10^{2}$ & $4 \cdot 10^{-2}$ & $-2.3$ \\
 $0.2$ & $1.43 \cdot 10^{9}$ & $4.1 \cdot 10^{-4}$ & $0.55$ & $21.0$ & $0.34$ & $-4.0$ & $5 \cdot 10^{2}$ & $-0.2$ & $4$ & $9.4 \cdot 10^{2}$ & $0.7$ & $-1$ \\
 $0.2$ & $4.56 \cdot 10^{14}$ & $5.3 \cdot 10^{-9}$ & $-1.4$ & $24.0$ & $2.3$ & $-3.1$ & $7 \cdot 10^{2}$ & $-8 \cdot 10^{-2}$ & $30$ & $8 \cdot 10^{2}$ & $9 \cdot 10^{-2}$ & $-1.9$ \\
 $0.2$ & $3.63 \cdot 10^{17}$ & $4.0 \cdot 10^{-11}$ & $-1.4$ & $22.0$ & $2.6$ & $-3.0$ & $8 \cdot 10^{2}$ & $-0.1$ & $20$ & $9.3 \cdot 10^{2}$ & $0.2$ & $-2$ \\

 $0.4$ & $75 \cdot 10^{6}$ & $1.4 \cdot 10^{-4}$ & $0.62$ & $87.0$ & $0.34$ & $-5.2$ & $6 \cdot 10^{2}$ & $-4 \cdot 10^{-2}$ & $70$ & $6 \cdot 10^{2}$ & $4 \cdot 10^{-2}$ & $-2.1$ \\
 $0.4$b & $1.43 \cdot 10^{9}$ & $2 \cdot 10^{-3}$ & $0.70$ & $3 \cdot 10^{2}$ & $4 \cdot 10^{-2}$ & $-20$ & $7 \cdot 10^{2}$ & $-2 \cdot 10^{-2}$ & $80$ & $8 \cdot 10^{2}$ & $3 \cdot 10^{-2}$ & $-2$ \\
 $0.4$ & $4.56 \cdot 10^{14}$ & $1.3 \cdot 10^{-9}$ & $-1.2$ & $50.0$ & $1.4$ & $-3.4$ & $3.4 \cdot 10^{2}$ & $-1.0$ & $0.4$ & $8.4 \cdot 10^{2}$ & $2$ & $-2.3$ \\
 $0.4$ & $3.63 \cdot 10^{17}$ & $1.1 \cdot 10^{-11}$ & $-1.2$ & $43.0$ & $1.6$ & $-3.1$ & $3.8 \cdot 10^{2}$ & $-1$ & $0.1$ & $1.0 \cdot 10^{3}$ & $2$ & $-2.3$ \\
\hline

 $0.8$a & $75 \cdot 10^{6}$ & $4 \cdot 10^{-2}$ & $15.0$ & $12.0$ & $5.6 \cdot 10^{-2}$ & $-2.9$ & $4.4 \cdot 10^{2}$ & $-3 \cdot 10^{2}$ & $-1.4$ & $8.9 \cdot 10^{2}$ & $20$ & $-2.4$ \\
 $0.8$ & $1.43 \cdot 10^{9}$ & $3 \cdot 10^{-5}$ & $8.3$ & $16.0$ & $1.7$ & $2.7$ & $3.4 \cdot 10^{2}$ & $-2.3$ & $5.9$ & $1.1 \cdot 10^{2}$ & $8.6 \cdot 10^{-2}$ & $-3.2$ \\
 $0.8$ & $4.56 \cdot 10^{14}$ & $3 \cdot 10^{-6}$ & $8.3$ & $8.6$ & $7.2 \cdot 10^{-2}$ & $-3.2$ & $4.0 \cdot 10^{2}$ & $-10$ & $-1.5$ & $9.0 \cdot 10^{2}$ & $2 \cdot 10^{1}$ & $-2.4$ \\
 $0.8$ & $3.63 \cdot 10^{17}$ & $3 \cdot 10^{-6}$ & $30$ & $1$ & $3 \cdot 10^{-2}$ & $-3.1$ & $4.3 \cdot 10^{2}$ & $-9$ & $-1.5$ & $1.0 \cdot 10^{3}$ & $20$ & $-2.4$ \\

 $\pi/2$a & $75 \cdot 10^{6}$ & $1.8 \cdot 10^{-5}$ & $8.0$ & $2.0 \cdot 10^{2}$ & $0.30$ & $-2.6$ & & & & & & \\
 $\pi/2$ & $1.43 \cdot 10^{9}$ & $3.8 \cdot 10^{-6}$ & $6.9$ & $1.6 \cdot 10^{2}$ & $0.25$ & $-2.6$ & & & & & & \\
 $\pi/2$ & $4.56 \cdot 10^{14}$ & $1.1 \cdot 10^{-10}$ & $3.4$ & $2.3 \cdot 10^{2}$ & $0.49$ & $-2.6$ & & & & & & \\
 $\pi/2$ & $3.63 \cdot 10^{17}$ & $6.9 \cdot 10^{-13}$ & $3.9$ & $2.3 \cdot 10^{2}$ & $0.39$ & $-2.5$ & & & & & & \\
\hline
\end{tabular}}
\caption{\footnotesize{Power law fit results for case A (top) \& B (bottom). The observer angle is in radians. Flux level $F_0$ is in mJy. The break times $t_{01}$, $t_{12}$ and $t_{23}$ are in days. The occasional lower case letters in the first column mark the following: a) Poor fit (meaning $\chi^2 / \text{dof} > 1$, with value up to $\sim 5$), caused by the complex shape of the initial rise of the light curve. b) Poor fit because the synchrotron break frequency $\nu_m$ passed through the observed band around ten days and the corresponding complication for the light curve was not taken into account (see also Figs. \ref{lightcurves_figure} and \ref{spectra_figure}). c) Good fit, albeit with the break times $t_{12}$ and $t_{23}$ swapped and the final slope rising, hindering an interpretation of $\beta_3$ as the final slope. When a fit was insensitive to a fit parameter and the resulting error on the fit parameter of the same order as the parameter itself, a single digit has been used. All light curves for case A run until 4,000 days observer time, with starting observer times of 1 day ($0.0$, $0.2$, $0.4$ rad), 15 days ($0.4$ rad) and 50 days ($\pi/2$ rad). For case B, the latest observer time covered is 3,200 days, with starting times of 1 day ($0.0$, $0.2$, $0.4$ rad), 6 days ($0.4$ rad) and 40 days ($\pi/2$ rad).}}
\label{characteristics_table}
\end{table}

\begin{table}
\tiny{\centering
\begin{tabular}{|ll||ll|lll|lll|lll|}
\hline
$\theta_{obs}$ & $\nu$ (Hz) & $F_0$ & $\beta_0$ & $t_{01}$ & $s_{01}$ & $\beta_1$ & $t_{12}$ & $s_{12}$ & $\beta_2$ & $t_{23}$ & $s_{23}$ & $\beta_3$ \\
\hline
 $0.0$ & $75 \cdot 10^{6}$ & $5.6 \cdot 10^{-3}$ & $0.64$ & $1.9 \cdot 10^{2}$ & $2.3$ & $-2.5$ & $1.8 \cdot 10^{3}$ & $-2$ & $1$ & $3 \cdot 10^{3}$ & $0.9$ & $-2.4$ \\
 $0.0$ &  $1.43 \cdot 10^{9}$  & $2.4 \cdot 10^{-2}$ & $0.41$ & $63.0$ & $1.5$ & $-3.0$ & $3 \cdot 10^{3}$ & $-0.2$ & $5$ & $4.2 \cdot 10^{3}$ & $0.6$ & $-1$ \\
 $0.0$ & $4.56 \cdot 10^{14}$ & $1.8 \cdot 10^{-6}$ & $-1.3$ & $61.0$ & $8.9$ & $-3.0$ & $3 \cdot 10^{3}$ & $-0.2$ & $6$ & $4.1 \cdot 10^{3}$ & $0.5$ & $-2$ \\
 $0.0$ & $3.63 \cdot 10^{17}$ & $6.4 \cdot 10^{-9}$ & $-1.3$ & $46.0$ & $4.0$ & $-2.9$ & $1.4 \cdot 10^{3}$ & $-0.51$ & $0.2$ & $4.6 \cdot 10^{3}$ & $8$ & $-1.1$ \\
 
$0.1$ & $75 \cdot 10^{6}$ & $5.5 \cdot 10^{-3}$ & $0.63$ & $1.9 \cdot 10^{2}$ & $1.9$ & $-2.5$ & $1.8 \cdot 10^{3}$ & $-2$ & $1$ & $3 \cdot 10^{3}$ & $0.9$ & $-2.4$ \\
$0.1$ &  $1.43 \cdot 10^{9}$ & $2.5 \cdot 10^{-2}$ & $0.42$ & $66.0$ & $0.98$ & $-3.1$ & $3 \cdot 10^{3}$ & $-0.2$ & $5$ & $4.1 \cdot 10^{3}$ & $0.5$ & $-2$ \\
$0.1$ & $4.56 \cdot 10^{14}$ & $1.3 \cdot 10^{-6}$ & $-1.3$ & $71.0$ & $3.5$ & $-3.0$ & $3 \cdot 10^{3}$ & $-0.2$ & $6$ & $3.9 \cdot 10^{3}$ & $0.4$ & $-2$ \\
$0.1$ & $3.63 \cdot 10^{17}$ & $5 \cdot 10^{-7}$ & $-1$ & $7$ & $0.4$ & $-1$ & $8.4 \cdot 10^{2}$ & $-1.2$ & $0.7$ & $56.0$ & $3.2$ & $-0.8$ \\

$0.2$ & $75 \cdot 10^{6}$ & $5.0 \cdot 10^{-3}$ & $0.58$ & $2.2 \cdot 10^{2}$ & $1.3$ & $-2.6$ & $1.7 \cdot 10^{3}$ & $-1$ & $2$ & $3 \cdot 10^{3}$ & $0.8$ & $-2.4$ \\
$0.2$ & $1.43 \cdot 10^{9}$ & $2.8 \cdot 10^{-2}$ & $0.42$ & $89.0$ & $0.47$ & $-3.6$ & $3 \cdot 10^{3}$ & $-0.2$ & $5$ & $4.1 \cdot 10^{3}$ & $0.6$ & $-2$ \\
$0.2$ & $4.56 \cdot 10^{14}$ & $5.0 \cdot 10^{-7}$ & $-1.4$ & $1.1 \cdot 10^{2}$ & $2.7$ & $-3.1$ & $2 \cdot 10^{3}$ & $-0.4$ & $4$ & $3.7 \cdot 10^{3}$ & $0.6$ & $-2.3$ \\
$0.2$ & $3.63 \cdot 10^{17}$ & $1.5 \cdot 10^{-9}$ & $-1.5$ & $93$ & $1$ & $-3.4$ & $2 \cdot 10^{3}$ & $-0.3$ & $1$ & $4.5 \cdot 10^{3}$ & $3$ & $-0.6$ \\

$0.4$ & $75 \cdot 10^{6}$ & $4.0 \cdot 10^{-3}$ & $0.58$ & $4.0 \cdot 10^{2}$ & $0.6$ & $-3$ & $1.7 \cdot 10^{3}$ & $-0.6$ & $4$ & $3 \cdot 10^{3}$ & $0.4$ & $-2.6$ \\
$0.4$ & $1.43 \cdot 10^{9}$ & $1.6 \cdot 10^{-2}$ & $0.41$ & $1.9 \cdot 10^{2}$ & $0.30$ & $-4.2$ & $1.6 \cdot 10^{3}$ & $-0.6$ & $0.9$ & $3.8 \cdot 10^{3}$ & $1$ & $-2.5$ \\
$0.4$ & $4.56 \cdot 10^{14}$ & $1.2 \cdot 10^{-7}$ & $-1.3$ & $2.3 \cdot 10^{2}$ & $2.4$ & $-3.2$ & $1.6 \cdot 10^{3}$ & $-1$ & $0.6$ & $3.9 \cdot 10^{3}$ & $1$ & $-2.5$ \\
$0.4$ & $3.63 \cdot 10^{17}$ & $2.5 \cdot 10^{-10}$ & $-1.4$ & $2.0 \cdot 10^{2}$ & $1$ & $-3.1$ & $2 \cdot 10^{3}$ & $-0.4$ & $2$ & $4.0 \cdot 10^{3}$ & $2$ & $-1$ \\

\hline

$0.8$a & $75 \cdot 10^{6}$ & $2 \cdot 10^{-2}$ & $5.9$ & $1.9 \cdot 10^{2}$ & $0.19$ & $-2.3$ & $1.9 \cdot 10^{3}$ & $-4$ & $2$ & $2.2 \cdot 10^{3}$ & $0.9$ & $-2.3$ \\
$0.8$a & $1.43 \cdot 10^{9}$ & $8.8 \cdot 10^{-3}$ & $8.5$ & $87.0$ & $0.20$ & $-2.1$ & $1.9 \cdot 10^{3}$ & $-3$ & $0.8$ & $2.9 \cdot 10^{3}$ & $0.5$ & $-2.9$ \\
$0.8$ & $4.56 \cdot 10^{14}$ & $2.4 \cdot 10^{-6}$ & $3.9$ & $99.0$ & $0.19$ & $-2.7$ & $1.9 \cdot 10^{3}$ & $-20$ & $-1.2$ & $4.3 \cdot 10^{3}$ & $8$ & $-2.4$ \\
$0.8$ & $3.63 \cdot 10^{17}$ & $4.4 \cdot 10^{-10}$ & $4.6$ & $65.0$ & $0.40$ & $-1.4$ & $1.7 \cdot 10^{3}$ & $-2$ & $2$ & $1.6 \cdot 10^{3}$ & $0.49$ & $-1.8$ \\

$\pi /2 $a & $75 \cdot 10^{6}$ & $1.6 \cdot 10^{-3}$ & $7.4$ & $1.0 \cdot 10^{3}$ & $0.34$ & $-2.6$ & & & & & & \\
$\pi/2$ & $1.43 \cdot 10^{9}$ & $4.8 \cdot 10^{-4}$ & $6.9$ & $7.7 \cdot 10^{2}$ & $0.23$ & $-2.8$ & & & & & & \\
$\pi / 2$ & $4.56 \cdot 10^{14}$ & $1.7 \cdot 10^{-8}$ & $3.8$ & $1.1 \cdot 10^{3}$ & $0.37$ & $-2.8$ & & & & & & \\
$\pi / 2$ & $3.63 \cdot 10^{17}$ & $1.5 \cdot 10^{-11}$ & $3.5$ & $1.0 \cdot 10^{3}$ & $0.43$ & $-2.0$ & & & & & & \\

\hline 
\hline

$0.0$d & $75 \cdot 10^{6}$  & $10 \cdot 10^{-15}$ & $6$ & $6$ & $-3 \cdot 10^{-2}$ & $-2$ & $11.0$ & $0.5$ & $-5$ & $1 \cdot 10^{3}$ & $1$ & $-7$ \\
$0.0$e & $1.43 \cdot 10^{9}$ & $0.16$ & $1.0$ & $1.1 \cdot 10^{2}$ & $0.2$ & $-9$ & $1.6 \cdot 10^{2}$ & $-0.4$ & $-0.3$ & $4.1 \cdot 10^{2}$ & $3$ & $-2.6$ \\
$0.0$ & $4.56 \cdot 10^{14}$  & $6 \cdot 10^{-5}$ & $-2.0$ & $5.2$ & $3.2$ & $-4.1$ & $2 \cdot 10^{4}$ & $-2 \cdot 10^{-2}$ & $20$ & $6.4 \cdot 10^{2}$ & $2$ & $10$ \\
$0.0$ & $3.63 \cdot 10^{17}$  & $5 \cdot 10^{-9}$ & $-2$ & $5.2$ & $2.6$ & $-4.5$ & $5 \cdot 10^{3}$ & $-2 \cdot 10^{-2}$ & $10$ & $6.2 \cdot 10^{2}$ & $2$ & $8$ \\

$0.1$ & $75 \cdot 10^{6}$ & $4 \cdot 10^{-5}$ & $1.2$ & $40$ & $0.2$ & $-8$ & $40$ & $-0.2$ & $1.1$ & $9.1 \cdot 10^{2}$ & $7$ & $6 \cdot 10^{-2}$ \\
$0.1$ & $1.43 \cdot 10^{9}$ & $0.15$ & $0.98$ & $1.1 \cdot 10^{2}$ & $0.16$ & $-8.7$ & $1.6 \cdot 10^{2}$ & $-0.5$ & $-0.7$ & $4.2 \cdot 10^{2}$ & $4$ & $-2.6$ \\
$0.1$ & $4.56 \cdot 10^{14}$ & $2 \cdot 10^{-4}$ & $-1.7$ & $5.4$ & $1.3$ & $-3.8$ & $7 \cdot 10^{3}$ & $-3 \cdot 10^{-2}$ & $10$ & $6.4 \cdot 10^{2}$ & $2$ & $10$ \\
$0.1$ & $3.63 \cdot 10^{17}$ & $2 \cdot 10^{-8}$ & $-2$ & $5.5$ & $1$ & $-4$ & $3 \cdot 10^{3}$ & $-3 \cdot 10^{-2}$ & $9$ & $6.1 \cdot 10^{2}$ & $2$ & $6$ \\

$0.2$d & $75 \cdot 10^{6}$ & $1 \cdot 10^{-5}$ & $1$ & $5.1$ & $10$ & $0.9$ & $2.5 \cdot 10^{5}$ & $2.4 \cdot 10^{2}$ & $-0.5$ & $2 \cdot 10^{3}$ & $0.6$ & $-0.5$ \\
$0.2$e & $1.43 \cdot 10^{9}$ & $9.8 \cdot 10^{-2}$ & $0.87$ & $1.1 \cdot 10^{2}$ & $0.2$ & $-9$ & $1.6 \cdot 10^{2}$ & $-0.5$ & $-0.7$ & $4.3 \cdot 10^{2}$ & $4$ & $-2.5$ \\
$0.2$ & $4.56 \cdot 10^{14}$ & $5.2 \cdot 10^{-5}$ & $-1.9$ & $11.0$ & $3.0$ & $-3.2$ & $4 \cdot 10^{2}$ & $-0.2$ & $3$ & $5.9 \cdot 10^{2}$ & $2$ & $-0.5$ \\
$0.2$ & $3.63 \cdot 10^{17}$ & $1.2 \cdot 10^{-8}$ & $-1.9$ & $12.0$ & $3$ & $-3.3$ & $2 \cdot 10^{2}$ & $-0.2$ & $1$ & $5.6 \cdot 10^{2}$ & $2$ & $-1$ \\

$0.4$d & $75 \cdot 10^{6}$ & $1 \cdot 10^{-6}$ & $1$ & $4$ & $-3$ & $1$ & $2 \cdot 10^{2}$ & $-10$ & $1$ & $1.0 \cdot 10^{3}$ & $10$ & $-7 \cdot 10^{-2}$ \\
$0.4$ & $1.43 \cdot 10^{9}$ & $5.9 \cdot 10^{-2}$ & $0.85$ & $1.3 \cdot 10^{2}$ & $0.23$ & $-7.3$ & $1.7 \cdot 10^{2}$ & $-1.2$ & $-2$ & $4.6 \cdot 10^{2}$ & $10$ & $-2.5$ \\
$0.4$ & $4.56 \cdot 10^{14}$ & $8 \cdot 10^{-6}$ & $-1.6$ & $30$ & $1$ & $-4$ & $1.6 \cdot 10^{2}$ & $-0.4$ & $0.6$ & $5.4 \cdot 10^{2}$ & $2$ & $-2$ \\
$0.4$ & $3.63 \cdot 10^{17}$ & $2 \cdot 10^{-9}$ & $-1.6$ & $40$ & $0.7$ & $-4$ & $1 \cdot 10^{2}$ & $-0.4$ & $0.2$ & $5.3 \cdot 10^{2}$ & $3$ & $-1.7$ \\

\hline

$0.8$ & $75 \cdot 10^{6}$ & $5.1 \cdot 10^{-6}$ & $3.5$ & $4.7$ & $1.6$ & $-1.6$ & $11.0$ & $-10$ & $1.5$ & $3 \cdot 10^{3}$ & $0.2$ & $-6$ \\
$0.8$ & $1.43 \cdot 10^{9}$ & $7 \cdot 10^{-4}$ & $6.9$ & $5.6$ & $0.80$ & $1.8$ & $68.0$ & $7$ & $0.2$ & $4 \cdot 10^{2}$ & $0.2$ & $-3.5$ \\
$0.8$ & $4.56 \cdot 10^{14}$ & $6 \cdot 10^{-4}$ & $9.3$ & $3.3$ & $0.11$ & $-2.0$ & $2.3 \cdot 10^{2}$ & $-60$ & $-1.0$ & $5.3 \cdot 10^{2}$ & $9$ & $-1.9$ \\
$0.8$ & $3.63 \cdot 10^{17}$ & $1 \cdot 10^{-7}$ & $9.1$ & $3.2$ & $0.12$ & $-2.0$ & $2.2 \cdot 10^{2}$ & $-60$ & $-1.0$ & $5.4 \cdot 10^{2}$ & $10$ & $-1.8$ \\

$\pi / 2$f & $75 \cdot 10^{6}$  & $5.8 \cdot 10^{-6}$ & $8.1$ & $28.0$ & $0.37$ & $1.4$ & & & & & & \\
   &   & $10 \cdot 10^{-4}$ & $1.5$ & $1.1 \cdot 10^{3}$ & $2$ & $-1$ & & & & & & \\
$\pi / 2$a & $1.43 \cdot 10^{9}$ & $6.6 \cdot 10^{-2}$ & $7.3$ & $1.1 \cdot 10^{2}$ & $0.29$ & $-2.6$ & & & & & & \\
$\pi / 2$  & $4.56 \cdot 10^{14}$ & $2.3 \cdot 10^{-6}$ & $3.8$ & $1.0 \cdot 10^{2}$ & $0.31$ & $-2.3$ & & & & & & \\
$\pi / 2$ & $3.63 \cdot 10^{17}$ & $3.7 \cdot 10^{-10}$ & $3.1$ & $1.1 \cdot 10^{2}$ & $0.41$ & $-2.0$ & & & & & & \\
\hline
\end{tabular}}
\caption{\footnotesize{Power law fit results for case C (top) \& D (bottom). The observer angle is in radians. Flux level $F_0$ is in mJy. The break times $t_{01}$, $t_{12}$ and $t_{23}$ are in days. Lower case letters in first column mark the following: a) Poor fit (meaning $\chi^2 / \text{dof} > 1$), caused by the complex shape of the initial rise of the light curve. d) Good fit, but the first two breaks represent complexities due to the combined effects of the jet structure and self-absorption, rather than jet break and rise of the counterjet. e) Good fit, but the double peak feature from the combined effect of jet structure and self-absorption is missed and the break times $t_{01}$ has no straightforward interpretation. f) There was such a clear break here in the rising phase that we fitted the curve in two parts: 15-600 days and 300-1500 days. When a fit was insensitive to a fit parameter and the resulting error on the fit parameter of the same order as the parameter itself, a single digit has been used. All light curves for case C run until 10,000 days observer time, with starting observer times of 2 days ($0.0$, $0.2$, $0.4$ rad), 30 days ($0.4$ rad) and 150 days ($\pi/2$ rad). For case D, the latest observer time covered is 1,500 days, with starting times of 1 day ($0.0$, $0.2$, $0.4$ rad), 3 days ($0.4$ rad) and 15 days ($\pi/2$ rad).}}
\label{characteristics2_table}
\end{table}

\end{document}